\documentclass[runningheads]{llncs}

\usepackage{cite}
\usepackage{amsmath,amssymb,amsfonts}
\usepackage{algorithmic}
\usepackage{graphicx}
\usepackage{textcomp}
\usepackage[dvipsnames]{xcolor}
\usepackage{booktabs}
\usepackage[hidelinks]{hyperref}
\usepackage{makecell}
\usepackage{multicol}
\usepackage{multirow}
\usepackage{url}
\usepackage{svg}
\usepackage{siunitx}
\usepackage[table]{pstricks}
\usepackage{pst-3d,pst-node}
\usepackage{soul}
\usepackage{caption}
\usepackage{subcaption}

\usepackage[framemethod=TikZ]{mdframed}

\definecolor{lightergray}{rgb}{0.97, 0.97, 0.97}
\definecolor{darkred}{rgb}{.6, 0, 0}
\definecolor{darkgreen}{rgb}{0,.4, 0}
\definecolor{darkblue}{rgb}{0, 0,.6}
\definecolor{darkergreen}{rgb}{0,.8,0}

\definecolor{mydarkblue}{RGB}{7, 85, 157}
\definecolor{mypurple}{RGB}{127, 40, 205}
\definecolor{mygreen}{RGB}{78, 178, 165}
\definecolor{lightblue}{RGB}{90, 161, 255}

\definecolor{lightseagreen}{rgb}{0.13, 0.7, 0.67}

\usepackage{colortbl}

  \presetkeys{todonotes}{
    fancyline,
    color=purple!30}{}

\usepackage{tabularx}
\usepackage{todonotes}
\usepackage[
  capitalise,
  nameinlink
]{cleveref}
\usepackage{hyphenat}
\usepackage{enumerate}
\usepackage{listings}

\usepackage{tikz}
\usetikzlibrary{positioning, matrix, arrows.meta, shapes.arrows, decorations.pathreplacing}
\usepackage{pgfplots}
\usepgfplotslibrary{statistics} 
\usepackage{xparse}
\NewDocumentCommand{\rot}{O{30} O{1em} m}{\makebox[#2][l]{\rotatebox{#1}{#3}}}%
 
\apptocmd{\UrlBreaks}{\do\f\do\m}{}{}
\begin{document}
\title{Reduce to the MACs - Privacy Friendly Generic Probe Requests}
%\titlerunning{Privacy Friendly Generic Probe Requests}
\author{Johanna Ansohn McDougall\inst{1}\orcidID{0000-0001-7457-2181} \and
Alessandro Brighente\inst{2}\orcidID{0000-0001-6138-2995} \and
Anne Kunstmann\inst{1}\orcidID{0009-0009-8990-9571}
 \and
Niklas Zapatka\inst{1}\orcidID{0009-0005-3841-1475}
 \and
Hannes Federrath\inst{1}}
\authorrunning{J. Ansohn McDougall et al.}

\institute{University of Hamburg, Hamburg, Germany \\
\email{\{johanna.ansohn.mcdougall, firstname.lastname\}@uni-hamburg.de} 
\and
University of Padova, Padua, Italy
\\ \email{alessandro.brighente@unipd.it}}

\maketitle

\begin{tikzpicture}[remember picture, overlay]
  \node[font=\sffamily\normalsize, yshift=-1cm, text centered, text width=\paperwidth, anchor=north west] at (current page.north west) {%
%This publication has been accepted at the 39th International Conference on ICT Systems Security and Privacy Protection (IFIP SEC 2024). The final publication and link will be available shortly.};
% The final publication is available at Springer via \url{https://doi.org/10.1007/978-3-031-09234-3_19}
%  };
This version of the
contribution has been accepted for publication, after peer review 
but is not the Version of Record and does not reflect post-acceptance improvements,
or any corrections. The Version of Record of this contribution
will be published in the proceedings of the 39th IFIP TC 11 International Conference (IFIP SEC 2024) and will be available online shortly. Use of this Accepted Version is subject to the
publisher’s Accepted Manuscript terms of use
\url{https://www.springernature.com/gp/open-research/policies/accepted-manuscriptterms}
};
\end{tikzpicture}
\vspace{-1cm}

%\begin{tikzpicture}[remember picture, overlay]
%  \node[font=\sffamily\normalsize, yshift=-1cm, text centered, text width=\paperwidth, anchor=north west] at (current page.north west) {%
%The final publication is available at Springer via %\url{https://doi.org/10.1007/978-3-031-09234-3_19}
% };
%\end{tikzpicture}

\begin{abstract}
Since the introduction of active discovery in Wi-Fi networks, users can be tracked via their probe requests. Although manufacturers typically try to conceal Media Access Control (MAC) addresses using MAC address randomisation, probe requests still contain Information Elements (IEs) that facilitate device identification.
This paper introduces \textit{generic probe requests}: By removing all unnecessary information from IEs, 
the requests become indistinguishable from one another,  %improving user privacy during active network discovery and 
letting single devices disappear in the largest possible anonymity set.

Conducting a comprehensive evaluation, we demonstrate that a large IE set contained within undirected probe requests does not necessarily imply fast connection establishment. Furthermore, we show that minimising IEs to nothing but Supported Rates would enable 82.55\% of the devices to share the same anonymity set. %Additionally, we ascertain that a generic MAC address not only enhances user anonymity during probing but also offers operational efficiency comparable to MAC address randomisation. 
Our contributions provide a significant advancement in the pursuit of robust privacy solutions for wireless networks, paving the way for more user anonymity and less surveillance in wireless communication ecosystems.

\end{abstract}


\section{Introduction}

Wireless communication enables seamless connectivity and data exchange among devices. Particularly the ubiquitous use of active discovery has raised serious concerns regarding user privacy: Probe requests are actively sent by devices attempting to join a network, and the first step required for connection establishment between a user's device and an Access Point (AP). %As they can contain a wealth of information \cite{PFP} and are probe to be used   
%and can be either directed or undirected. When directed, they contain one or more entries of their a Preferred Network List (PNL): a list of all known which may include identifying or private information . Yet even undirected probe requests that do not contain a PNL can be used to track a device due to the contained MAC address. To mitigate this, a randomised MAC address is used instead of the real hardware MAC address. 
%Manufacturers use various approaches in order to implement MAC address randomisation \cite{Vanhoef-RandomisationNotEnough}, as well as researchers propose their original schemes \cite{he2022self, gomez2022evolution}. While its standardisation was proposed in February 2022, it is, as of today, still in its draft stage \cite{MAR-rfc}. The manufactures therefore implement randomisation as they see fit. One approach is to keep unchanged the first three bytes of the MAC address, also known as the Organizationally Unique Identifier (OUI), and randomise only the last three bytes \cite{Vanhoef-RandomisationNotEnough, martin2017study}. However, this allows attackers to still infer information on devices, and possibly track and trace them. 
%While this technique protects the device from disclosing its real MAC address, % is implemented in a way that prevents attackers from using the MAC address as an identifier for tracking, 
Albeit modern devices typically use MAC address randomisation and omit to send known Service Set Identifiers (SSIDs) unless searching for hidden networks \cite{PFP},
the information contained within the IE of the probe request still serves as a device fingerprint \cite{TalkativePhones, Vanhoef-RandomisationNotEnough, martin2017study, ProbeReqIdentification, uras-ai, pintor-ie-fingerprinting, rutermann2019know, tan2021efficient}. This fingerprint can be used to track device users, approximate the amount of people in a specified area and even trilaterate and thereby locate the signal origin up to an accuracy of 1.5\,m \cite{PFP}.

In response to these limitations, this paper introduces \textit{generic probe requests}: 
We propose the reduction of IE content to the bare minimum. % and the consistent use of a single MAC address for all devices over all probe requests.  
We study the implications of generic probe requests for functionality, privacy, device identification, and tracking prevention. 
%offering a comparable performance to established schemes like MAC address randomisation. 
To this end, we contribute the following:

\begin{itemize}
    \item We analyse the IE of probe requests and determine the minimal information required to receive probe responses. Our results show that the SSID and Supported Rates field are sufficient to receive valid probe responses.
    \item We analyse the impact of a reduced IE on functionality, privacy and security. % necessity of IE with respect to connection establishment and exchange of capabilities.
    \item We propose Generic Probe Requests and show that they provide a sufficiently large anonymity set, encompassing \num{82.55}\% of the probe requests. Thanks to the reduced content of their IEs, Generic Probe Requests resist correlation attacks undermining user privacy.
    %\item To defend against attacks targeting MAC address randomisation, we propose and implement the Generic Address Scheme. Our results show that MAC address randomisation can be replaced by a single fixed MAC address for all devices. We further show that the time required for connection establishment using our scheme is comparable to that of MAC address randomisation, but providing higher privacy guarantees.
\end{itemize}

Our results show that Generic Probe Requests %and generic addresses 
make devices as indistinguishable as possible, defending users from attacks targeting device fingerprinting via the IE content while simultaneously having no impact on the actual connection establishment, since they are only used during active discovery. 
%This technique . 
While various publications have proposed to reduce IE content and randomise sequence numbers \cite{Vanhoef-RandomisationNotEnough, martin2017study, matteDefeatingMACAddress2016}, to the best of our knowledge, we are the first to  study the implications of minimising and unifying probe requests to the maximum.

This paper is structured as follows: In \cref{sec: Background}, we provide a background on MAC addresses, probe requests and  connection establishment in Wi-Fi networks. Additionally, we introduce the Time-to-Traffic metric.
We present Related Work in \cref{sec: Related Work} and the attacker model in \cref{sec: Attacker Model}. \Cref{sec: generic probe requests} introduces the concept of generic probe requests and analyses its impact on functionality, privacy and security. %We subsequently analyse the impact a reduced IE set has on existing attacks. % \cref{sec: generic address scheme} presents the implementation and test setup of generic addresses.
Finally, \cref{sec: Conclusion} concludes the paper.

\section{Background}\label{sec: Background}

This section gives a background on MAC addresses, probe requests and Wi-Fi connection establishment. 
In addition to that, we introduce the Time-to-Traffic metric that is used in our experiments.

\subsection{MAC Addresses}\label{sec:Mac addresses and collisions}

A MAC address is a 48-bit network address used in Wi-Fi networks \cite{WLAN-spec}.
It serves to identify frame senders and receivers on the data link layer.
%If two or more nodes within a network share the same MAC address, a MAC address collision occurs.
%Affected nodes can no longer be distinguished on the data link layer; consequently, a frame intended for a single destination will also be received by nodes with colliding addresses. 
Every Network Interface Controller (NIC) is assigned % a Universally Administered Address (UAA).
a fixed and globally unique Universally Administered Address (UAA)  by its manufacturer. 
The first three bytes,  the Organisationally Unique Identifier (OUI), identify the manufacturer. The last three bytes are assigned by the manufacturer.
The UAA can be substituted by a Locally Administered Address (LAA).
%In this case, network-wide uniqueness must be enforced manually.
UAAs and LAAs are distinguished by the U/L-bit contained in every MAC address, which is the second-least significant bit of the most significant byte of a MAC address \cite{IEEE802}.
An example of LAAs are randomised MAC addresses, used to conceal the UAA while sending probe requests to prevent tracking via the MAC address. %When MAC address randomisation is used, the MAC address is typically used for a whole burst of packets, i.e.,
%Different devices use different randomisation strategies: 
%Some devices, such as the iPhone SE models used in \cref{sec: generic probe requests}, re-randomise their address after every burst. 
%Other devices such as the DWA-171 antenna re-randomise it less frequently, after several bursts from the same address. 

\subsection{Probe Requests}\label{sec: probe requests}

A mobile device can identify known networks via active or passive scanning. 
%To establish a connection between a client and an AP, two means of network discovery can be used: active or passive scanning. 
During passive scanning, APs broadcast beacon frames containing their SSID every 102.4\,ms. Clients scanning the beacons can initiate the connection establishment upon recognising a known SSID \cite{GoovaertsPassiveNotActive}.
The alternative, active scanning, is more commonly used because of its reduced overhead \cite{GoovaertsPassiveNotActive}.
Here, clients broadcast \textit{bursts} of probe requests: a set of bundled requests sent within a short time and transmitted over several channels  \cite{PFP}. Modern devices typically send \textit{undirected} probe requests containing an empty SSID tag. An AP receiving such a probe request responds with a probe response containing its SSID.
\textit{Directed} probe requests contain SSIDs of known networks. They are required to locate hidden networks, which do not advertise via beacons and only respond to directed probe requests. Devices running outdated Operating Systems (OSes), or ones misconfigured by their users, may also include SSIDs in their probe requests \cite{PFP}.

The transmission of probe requests is initiated by the MAC sublayer management entity (MLME), which invokes layer management functions. The MLME initiates a scan via the MLME-SCAN.request primitive, which also determines the content of the probe request \cite{WLAN-spec}. This content can the be surveyed in the 
IE. While a large range of parameters \textit{can} be transmitted via the IE \cite[Section 6.3.3.2.2]{WLAN-spec}, devices typically transmit a select few IE tags. These can include, but are not limited to the Supported Rates, Extended Supported Rates and the transmission channel (DS Parameter). Other common fields are High Throughput (HT), Very High Throughput (VHT) and High Efficiency (HE) Capabilities, advertising support for the IEEE 802.11n,  802.11ac and 802.11ax standard respective. Extended Capabilities declare support for additional features beyond HT and VHT capabilities.  The Interworking field enables seamless connectivity in heterogeneous network environments (IEEE 802.11u). Another common field contains vendor specific information.

\subsection{Connection Establishment in Wi-Fi Networks}\label{sec: Connection Establishment in Wi-Fi Networks}

Wireless local area networks (WLANs) are based on the IEEE 802.11 standard \cite{WLAN-spec}.     
%Mobile clients connect to the WLAN via Access Points (APs) acting as base stations. 
Each WLAN is identified by its SSID, and mobile devices can identify known networks via network discovery (cf. \cref{sec: probe requests}). Subsequent connection establishment with an AP is done in several steps: first, IEEE 802.11 authentication is performed to grant a client access to the network. This encompasses two authentication and two association frames, in which the client and AP are paired and communication parameters and standard extensions negotiated.
A successful association enables data transfer of frames of higher layers. Subsequently, WPA, WPA2 and WPA3 use Robust Security Network Association (RSNA), a suite of protocols for authentication and encryption.
Upon successful completion, WLAN access is granted, and encrypted data frames can be exchanged.

While the term WLAN describes the standard, the Wi-Fi Alliance instead promotes the use of the term Wi-Fi, a trademark protecting all products certified to their Wi-Fi interoperability \cite{wiki-wifi}. 
Since the term Wi-Fi is commonly used in anglophone publications, we adhere to this de-facto standard in this paper.

\subsection{Time-to-Traffic}\label{sec: TtT}

\begin{figure}[t]
        \centering
       \frame{\includegraphics[width=\textwidth,trim=1mm 0 0 2mm, clip]{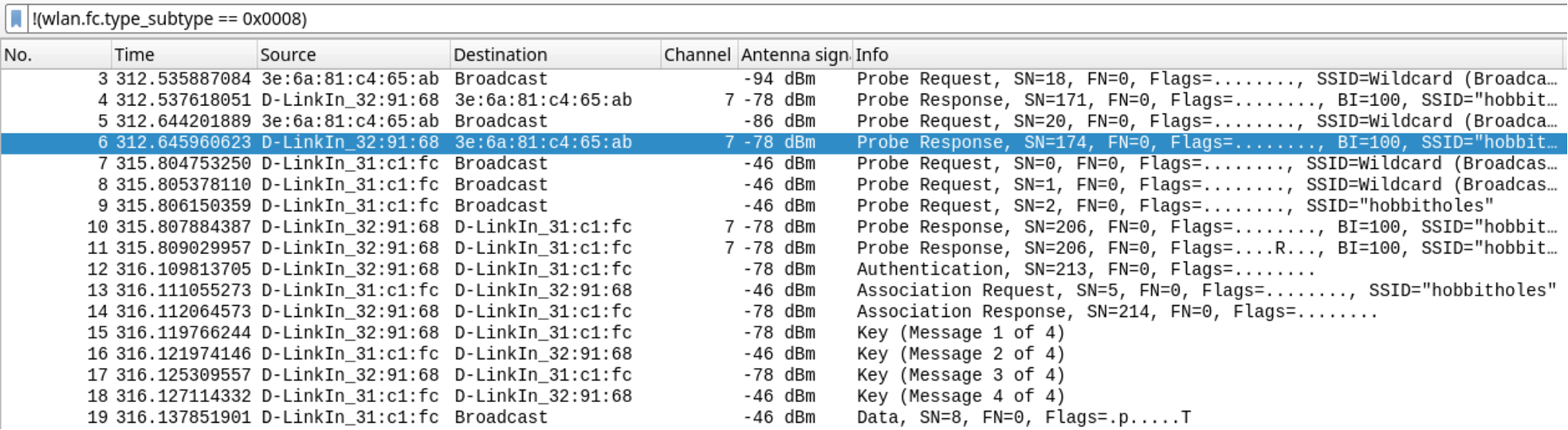}}
        \caption{Measurement points in the TtT-metric. Packet 6 is the starting point, the first probe response to the last probe request sent via a randomised address. The endpoint of the metric is the beginning of data transmission in packet 19.}
        \label{fig:client-ttt}
\end{figure}
We introduce the Time-to-Traffic (TtT) metric for measuring the duration of the connection process. With it, we measure the time between the last probe response received before the MAC address changes to the LAA, and the first data frame exchanged after connection establishment. Its duration can be observed in \cref{fig:client-ttt}: Packet 6, marked in blue, is the starting point of the TtT. 
The endpoint of the measurement is the first data frame, packet 19. This can be explained as follows: a device probing for nearby networks usually uses a randomised MAC address. The probe requests in this stage are typically undirected and therefore do not contain an SSID. Once a device identifies a known network via a probe response, it switches its MAC address to its UAA or a fixed LAA per network. %, or, as for example in the case of the iPhone SE used in \cref{sec: generic probe requests}, to. 
Typically, further probes are transmitted via this address, sometimes first undirected and then directed, sometimes either of the two. Subsequently,  connection establishment is initiated. Some devices %like the DWA-171 adaptor (cf. \cref{sec: generic probe requests}) 
mostly omit these directed or undirected probe requests using the LAA, and immediately initialise the connection after receiving a probe response from a known network. To receive comparable results, we therefore choose to measure the TtT beginning with the last probe response to the randomised MAC address.

\section{Related Work}\label{sec: Related Work}

%By monitoring probe requests sent by nearby devices, an eavesdropper can triangulate their origin.
To track a mobile device over a longer period of time, it is necessary to apply de-randomisation techniques to probe requests to correlate the bursts originating from different MAC addresses with a single device.
The research on this topic can be split into two competing perspectives: attacks, and countermeasures.

\subsection{Attacks}

Linking bursts to devices requires a unique pattern in the requests, e.g. the globally unique UAA \cite{bernardosWiFiInternetConnectivity2015}.
When a device uses a regularly changing LAA, other fingerprinting attacks are possible.
%In these cases, %the MAC address is not an identifier but an additional source of information, while 
The main focus then lies in fingerprinting using IEs and sequence numbers.
After the theoretical risk was noticed \cite{TalkativePhones}, a 
first empirical assessment was provided \cite{Vanhoef-RandomisationNotEnough}.
This attack was improved by taking the MAC address into account \cite{martin2017study}.
Subsequently, other IE fields were shown to be usable for fingerprinting devices \cite{Robyns2017Noncooperative8M,ProbeReqIdentification,He2023SelfSupervisedAO}.  
Other approaches to defeat MAC address randomisation model probe request frame association in a flow network and use minimum-cost flow optimisation \cite{tan2021efficient}, or combine IE fingerprinting with clustering approaches~\cite{uras-ai, pintor-ie-fingerprinting}.%, with Uras et al. \cite{uras-ai} reaching an accuracy between 65.2\% and 91.3\%, and Pintor et al. \cite{pintor-ie-fingerprinting} up to 92\%.

An additional source of information for device fingerprinting are timing attacks:
Pattern in request transmission times were shown to be specific to the device driver \cite{beaconsOnly}.
This attack was subsequently extended to single out individual devices \cite{matteDefeatingMACAddress2016}.
Using a timing attack, it is possible to identify a device's hardware by measuring channel switch times during scanning \cite{GoovaertsPassiveNotActive}.

\subsection{Countermeasures}

Albeit fingerprint techniques are improved continuously, manufacturers are slow to deploy countermeasures. The solutions are mainly academic, e.g. to rely solely on passive discovery, which decelerates the connection establishment by only 0.6 seconds \cite{waltari2018informationleak}. This can be improved by accelerating passive scanning \cite{GoovaertsPassiveNotActive}. 
While this is a valid approach in most cases, hidden networks can only be reached via directed probe requests. By transmitting the SSID as a hash, combined with the current MAC address and sequence number of the packet, the use of hidden networks is possible in a privacy friendly  manner \cite{PFP}.   

%While this is a valid means to protect the cleartext SSID and prevents tracking via the SSID fingerprint, fingerprinting the IE is the most prevalent way of circumventing anti-tracking techniques and is not countered sufficiently in existing publications.

%Preventing probe request tracking within active scanning requires for probe requests to no longer contain identifying information.
Improvements to active scanning include the use of 
MAC address randomisation, avoiding unique MAC addresses by using a regularly changing, randomised LAA \cite{gruteser2003EnhancingLocationPrivacy}.
An alternative is MAC address masquerading, where the own UAA is replaced by the MAC address of a nearby device \cite{beaconsOnly}.
Fingerprinting could be further hampered by minimising transmitted IEs and randomising sequence numbers \cite{Vanhoef-RandomisationNotEnough,martin2017study}, and introducing randomness into the sending pattern of single packets and bursts \cite{matteDefeatingMACAddress2016}. The use of cryptographic measures for content protection of probe requests would create an overhead of 2 seconds for the key exchange and 0.5 seconds for the transmitted probe request \cite{ProbeReqIdentification}. %Such a large overhead induces the question whether completely relinquishing the use of probe requests in favour of passive discovery might be more efficient. 
%While the additional time required might still be feasible, the complete protocol change of both client and AP is unlikely to be implemented.  
To assess the countermeasures deployed in mobile devices, a comparison of 160 mobile devices manufactured between 2012 and 2020 revealed that MAC address randomisation has been widely adopted, and sequence number randomisation is emerging.
However, IE fingerprinting is still possible \cite{MARAndWhenItSucceeds}.

Our proposal to employ generic probe requests enhances these countermeasures:
The IE minimisation maximise the anonymity set size and ensures that tracking via the content of probe requests is impeded.

\section{Attacker Model}\label{sec: Attacker Model}

We consider a passive attacker. Their objective is to track the movements of individuals within a public area, such as a campus or shopping centre.
This is accomplished by eavesdropping on the probe request bursts emitted by smartphones during active scanning.
The adversary has installed a sufficient number of Wi-Fi receivers in the area to triangulate their targets' positions.

We assume that the attacker can link probe request bursts to individual devices by exploiting unique patterns, e.g. the UAA. If a device uses MAC address randomisation, a pattern may still be derived by fingerprinting the associated IEs.
If two probe requests exhibit the same patterns, the attacker assumes that they originate from the same device.
We assume the attacker is thereby capable of associating probe requests to their sending devices and can track single users. 

%We assume that the attacker identifies bursts solely based on a locally unique MAC address.
%If sequence numbers or IEs are sufficiently identifying to link individual probe requests to the same device, other countermeasures such as sequence number randomisation or IE removal must be applied.

% We consider a passive attacker who wants to infer the movement of people in an area by monitoring their device's probe requests. We assume that the attacker has a sufficiently large number of distributed receivers with which she can monitor an area, e.g. a large shop or mall or a university campus.

%\todo[inline]{Niklas: Is this sufficient? Do we need a threat model?}

%\subsection{Threat Model}

%Our threat model considers the following attacks:
\section{Generic Probe Requests}\label{sec: generic probe requests}

Various attacks to track devices despite MAC address randomisation utilise IE-fingerprinting techniques \cite{Vanhoef-RandomisationNotEnough, martin2017study, uras-ai, pintor-ie-fingerprinting, ProbeReqIdentification}. Our approach to counter them hence lies in reducing the content of the IE to the bare minimum. 
To gauge how far the content of a probe request can be reduced, we transmit probe requests  via a USB Wi-Fi adaptor using Scapy\footnote{\url{https://scapy.net/}}, an interactive program for packet manipulation in Python. Two more Wi-Fi adaptors are used to monitor the traffic via Wireshark on the non-overlapping channels 1 and 6 of the 2.4 GHz bands, which are the most frequently used channels in the vicinity of our experimental setup. %We capture and replay probe requests sent by nearby devices in Scapy. 
The script used to send probe requests can be found on  GitHub\footnote{\url{https://Github.com/heddha/scapy-probe-requests}}.

\newcounter{observ}[section]\setcounter{observ}{0}
\renewcommand{\theobserv}{\arabic{observ}}
\newenvironment{observ}[2][]{%
\refstepcounter{observ}%
\mdfsetup{linecolor=mypurple!20,%
linewidth=2pt,topline=true,%
frametitleaboveskip=\dimexpr-\ht\strutbox\relax
}
\begin{mdframed}[]\relax%
\textbf{Observation~\theobserv: }\label{#2}}{\end{mdframed}}

\begin{observ}{observ1}
By removing more and more fields from the IE of the probe request, we determine that the SSID and Supported Rates field are the only ones necessary for probe requests. Neither of these fields is required to contain legitimate information in order to allow for appropriate reception of probe responses. 
\end{observ}

%    \item Regardless of repeated usage of the same source address as well as the number of requests sent, the APs in our vicinity reply to probe requests.
%\end{enumerate}  

Since the content of the IE is typically used for tracking, we propose to reduce it to the bare minimum established in this experiment.  % we identified as useful to receive probe responses. %Additionally, to make probe requests as generic as possible, we propose to send all probe requests from the same MAC address while probing, and only switching to the LAA once connection establishment is attempted.
To evaluate the feasibility of this approach, we subsequently evaluate the time required for connection establishment depending on the content of the IE to determine whether a large IE causes faster connection establishment. % in \cref{sec: ie size test}.% We then use a large data set to evaluate the impact the reduction of IE content has on the anonymity set size in \cref{Anonymity Set Determination}. Subsequently, we evaluate the impact this reduction has on previously introduced attacks in \cref{sec: Resistance to IE-Concerned Attacks}. % The second experiment determines whether generic addresses can be used for probing and whether this impacts performance in \cref{sec: generic address scheme}.

\subsection{IE Content Analysis}\label{sec: ie size test}

In theory, IEs in Probe Requests serve clients to communicate their capabilities to nearby APs. This information helps the APs to respond appropriately and efficiently, ensuring that the client can connect to the network in a manner aligning with its requirements and the network's capabilities.
To estimate whether transmitting capabilities using a complex IE field accelerates the connection establishment, we compare the connection establishment times of five different devices and correlate them with the size of their IE. The devices used in our tests are a mobile phone (iPhone SE 2020 running iOS 17), a Wi-Fi dongle (D-Link DWA-171 Nano Wi-Fi USB Adaptor), a laptop running an Intel AC 8265 wireless card, a single-board computer (Raspberry Pi Model 3B+), and a tablet (iPad Pro running iPadOS 15). They represent typical household appliances.%: a mobile phone, a tablet, a laptop, a Wi-Fi dongle and a single-board computer, e.g. used in an IoT device.

%adapt our Time-to-Traffic (TtT) metric: The starting point of the measurement is the time of the first probe response received directly after the last probe request was sent from a randomised MAC address. The end point is the first data frame. This can be explained as follows: a devices probing for nearby networks typically uses a randomised MAC address. The probe requests in the stage are typically undirected and therefore do not contain an SSID. Once a device recognises a known SSID in a probe response, it switches it's MAC address, either to its burned-in address, or, as for example in the case of the Apple device, to a fixed MAC address per network. With this address, it sometimes probes once or twice more using first an undirected, then a directed probe request, and then initiates the connection establishment. Some devices like the DWA-171 adapter mostly omit these directed or undirected probe requests using the LAA, and immediately initialise the connection after receiving a probe response from a known network. To receive comparable results, we therefore chose to measure the TtT beginning with the last probe response to the randomised MAC address.

The tests are conducted as follows: We connect our devices to an access point, which we turn on at non-fixed intervals. We measure how long it takes the devices to reconnect to the AP. To measure whether a complex information element causes quicker connection establishment, we utilise the TtT metric introduced in \cref{sec: TtT}. We conduct the tests 7 times per device.

\begin{table}[tb]
\begin{scriptsize}
    \caption{Comparison of the Information Element and the respective Time-to-Traffic of probe requests of five different devices.}
    \label{tab:ie-comparisons}
    \centering
      \setlength{\tabcolsep}{1.6ex}
  \begin{tabular}{l*{13}{c}}
%    \begin{tabular}{c|c|c|c}
     & 
    \rot{Supported Rates (bytes)} &
    \rot{Ext. Supported Rates (bytes)} &
    \rot{DS Parameter Set (bytes)} & %: Current Channel} &
    \rot{HT Capabilities (bytes)} &
    \rot{Vendor Specific (bytes)} &
    \rot{Extended Capabilities (bytes)} &
    \rot{HE Capabilities (bytes)} &
    \rot{Interworking (bytes)} &
    \rot{Frame Size undirected (bytes)} &
    \rot{Frame Size directed (bytes)} &
    \rot{ Average TtT (seconds)} &
    \rot{Median} &
    \rot{Std. Deviation}
 \\
        \toprule
        DWA-171 & 8 & 4 & - & -& -& -&- & -& 68 & 79 & \num{3.71} & \num{3.69} & \num{0.16} \\
        Intel 8265 & 8 & 4 & 1 & 26 & 7 & - & - & - & 108 &  277 & \num{1.92} & \num{1.67} & \num{0.73}\\
        iPhone SE & 4 & 8 & 1 & 26 & - & 8 & 27 & - & 139 &  184 & \num{2.77} & \num{2.41} & \num{0.98}\\
        iPad Pro & 4 & 8 & 1 & 26 & 28 &  8 & - & 7 & 152 & 163 & \num{2.54} & \num{2.8} & \num{1.17}\\
        Raspberry Pi & 4 & 8 & 1 & 26 & 131 &  - & - & - & 236 & 247 & \num{2.73} & \num{1.86} & \num{2.26}\\
        \bottomrule
    \end{tabular}
\end{scriptsize}
\end{table}

The results of the experiment can be seen in \cref{tab:ie-comparisons}.
All devices transmit a minimum of Supported Rates and Extended Supported rates. The probe requests of the \textit{DWA-171 adaptor} have a total frame size of 68 bytes, 12 bytes of which made up by Supported Rates and Extended Supported Rates. The TtT required is \num{3.71} seconds on average. The \textit{Intel 8265} laptop transmits undirected probe requests with a total frame size of 108 bytes. The TtT is \num{1.92} seconds on average, and the IE additionally contains channel information, HT Capabilities, and vendor-specific information. The frame size of the \textit{iPhone} is 139 bytes, and its TtT \num{2.77} seconds. In addition to the channel and HT Capabilities, it transmits 35 bytes of Extended Capabilities and HE Capabilities. The \textit{iPad Pro} transmits 152 bytes long probe request frames and has a TtT of 2.54 seconds. 
%It additionally transmits the channel, HT Capabilities, vendor-specific elements concerning Apple, Microsoft, and Broadcom and an Interworking tag, as well as Extended Capabilities.     
The largest IE body is contained in the undirected probe requests of the \textit{Raspberry Pi}, with 236 bytes and a  TtT of \num{2.73} seconds. %It additionally transmits DS Parameters, HT Capabilities, and three vendor specific tags, concerning Broadcom and P2P, and also a 105 byte long tag concerning WPS.  

As this comparison shows, the Intel 8265 card is by far the fastest despite the second-smallest frame size. This is also represented by its median of 1.67. The Intel 8265's standard deviation is the second smallest of all with 0.73. The iPad follows as the second-fastest, however, its standard deviation of 1.17 is higher than both the iPhone or the Raspberry Pi (with the largest frame body). The DWA-171 antenna is the slowest to establish a connection and also has the smallest frame size. With a median of 3.69, it has the smallest standard deviation of only 0.16. % the Raspberry Pi had by far the largest frame body, but only ranged in mid-field with respect to connection establishment.
The results show that there does not appear to be a direct correlation between the amount of information transmitted in undirected probe requests and the speed of the connection establishment after the device discovery. The underlying reasons for fast connection establishment can likely be found in optimised implementations, and the very slow connection establishment of the DWA-171 might be attributed to the slow data transmission via a USB 2.0 Type A wireless dongle. %Despite this, there does not appear to be a direct correlation between fast connection establishment and the amount of information transmitted in the IE during undirected probing.

On the other hand, \cref{tab:ie-comparisons} also allows for a comparison of the frame sizes of \textit{directed} probe requests: %, the frame sizes of the Intel 8265 and the Raspberry Pi were considerably larger than the iPhone, the iPad and the DWA-171 antenna: 
Directly before connection establishment between client and AP as well as after the client has switched to the LAA, it often sends one or more directed probe requests.  
A directed probe request from the DWA-171, the iPad, and the Raspberry Pi contains no additional information besides the SSID.
The iPhone transmits 184 instead of 139 bytes in its directed probe request, with the additional bytes concerning three vendor-specific tags and the SSID.  
The Intel 8265 laptop transmits a directed probe request of 277 bytes, with the additional information concerning Extended Capabilities, Mesh ID, FILS request parameters, and three vendor-specific tags containing information on WPS, P2P, and multi-band operation. 
While this does not necessarily have to correlate, comparing the connection establishment with respect to the IE size of directed probe requests could be interesting in future work.

\subsection{Proposition: Minimisation of IE content}
In order to not impede any significance that directed probe requests might have on communication establishment,  we propose \textit{generic probe requests, in which the IE content of undirected probe requests is reduced to the bare minimum}: By modifying the IE to only contain the Supported Rates and (empty) SSID fields, user privacy can be increased immensely, while simultaneously satisfying only the minimum requirements for legitimate IE content. The modifications only have to be applied to undirected probe requests as the majority of transmitted probe requests and are the core element of the research on deanonymisation of probe requests. Protecting them is essential to assure user privacy. 
The reason for limiting the protection to undirected probe requests lies in the nature of \textit{directed} probe requests, which  are only transmitted in three scenarios: 
\begin{enumerate}[(i)]
    \item Directly before connection establishment, 
    \item during an existing connection to ensure the connection is maintained with the AP with the strongest signal and 
    \item while searching for hidden networks.
\end{enumerate}
% (i) Directly before connection establishment, 
% (ii) during an existing connection to ensure the connection is maintained with the AP with the strongest signal and 
% (iii) while searching for hidden networks.

In the first two cases, the directed probe requests are sent using the LAA, which is typically either stable for an extended period of time per network, or unchanging \cite{PFP}. The transmission of the LAA and that of the SSID serve as a good enough identifier to enable trivial device tracking, which makes the protection of IE content irrelevant in this scenario. %Our experiments showed that more information contained in the IE do not necessarily correlate with the fastest connection establishment. We therefore argue that omitting all fields apart from Supported Rates and the SSID from probe requests is a valid option to achieve a higher privacy gain. 
Limiting the content of undirected probe requests has no impact on connection establishment or the connection itself: Both adhere to standards that do not rely on probe requests, using the established methods explained in \cref{sec: Connection Establishment in Wi-Fi Networks}. 
%The robustness of the connection and data transfer, connection speed, functionality and reliability are therefore not influenced by the use of generic probe requests. 

In the following, we investigate the impact a reduced IE content has on the exchange of capabilities with the AP.

\subsection{Impact on Functionality}

%The literature on probe requests typically describes the functionality of the Information Elements as required to transmit the functionality support and capabilities of a device \cite{pintor-ie-fingerprinting, Vanhoef-RandomisationNotEnough}. 
Connection establishment as described in \cref{sec: Connection Establishment in Wi-Fi Networks} requires four stages: (1) active or passive device discovery, (2) authentication, (3) association and (4)~RSNA. 
\begin{observ}
    CConnection establishment is also possible when using passive discovery only. Since passive discovery results in a connection just as efficient and stable as when using active discovery \cite{GoovaertsPassiveNotActive, waltari2018informationleak}, the capabilities that define the connection must be exchanged in another manner.
\end{observ}
When studying the IEEE Wi-Fi standard \cite[Fig. 4-30]{WLAN-spec}, it becomes apparent that when the AP sends probe responses, as well as when the client sends the IEEE Std. 802.11 association request, ''security parameters`` are transmitted. 
%Both the association request and the probe request are transmitted by the mobile device, while the probe response is transmitted by the AP. 
A comparison of the fields contained in association requests \cite[Table 9-34]{WLAN-spec} and probe requests \cite[Table 9-38]{WLAN-spec} reveals that association requests and probe requests contain many overlapping fields, but while association requests contain 46 elements, probe requests contain only 34 elements. Out of these 34 elements, 15 are not present in association requests. They can be observed in \cref{tab:comparison-ie-fields}. %, the exact :  Request, DSSS Parameters, SSID List, Channel Usage, Mesh ID, Estimated Service Parameters Inbound, Extended Request, FILS Request Parameters, AP-CSN, Change Sequence, S1G Relay Discovery, PV1 Probe Response Option, Vendor Specific Request, Cluster Probe and Estimated Service Parameters Outbound.   

% \begin{table}[t]
% \begin{scriptsize}
%     \caption{Comparison of the Information Element items in different types of requests.}
%     \label{tab:ie-items}
%     \centering
%       \setlength{\tabcolsep}{2ex}
%   \begin{tabular}{l*{3}{l}}
% %    \begin{tabular}{c|c|c|c}
%       & Association Requests & Probe Requests & Probe Responses\\
%         \toprule
%         No. of IE items & 46 & 34 & ?\\[2ex]
%         Exclusive IE items  & - & \checkmark & -\\
%         \quad AP-CSN & - & \checkmark & \checkmark\\
%         \quad Change Sequence & - & \checkmark & \checkmark\\
%         \quad Channel Usage & - & \checkmark & \checkmark\\
%         \quad Cluster Probe & - & \checkmark & -\\
%         \quad DSSS Parameters & - & \checkmark & \checkmark\\
%         \quad Estimated Service Parameters Inbound & - & \checkmark & \checkmark\\
%         \quad Estimated Services Parameters Outbound & - & \checkmark & \checkmark\\
%         \quad Extended Request & - & \checkmark & -\\
%         \quad FILS Request Parameters & - & \checkmark & -\\
%         \quad Mesh ID & - & \checkmark & \checkmark\\
%         \quad PV1 Prove Response Option & - & \checkmark & -\\
%         \quad Request & - & \checkmark & -\\
%         \quad S1G Relay Discovery & - & \checkmark & \checkmark\\
%         \quad SSID List & - & \checkmark & -\\
%         \quad Vendor Specific Request & - & \checkmark & -\\ 
%         \bottomrule
%     \end{tabular}
% \end{scriptsize}
% \end{table}

\begin{table}
\caption{The fields present in probe requests, but \textit{not} in association requests, and their existence or substitution by other fields in association requests and probe responses. }
\label{tab:comparison-ie-fields}
\begin{scriptsize}
    \newcommand\y{\checkmark}
      \setlength{\tabcolsep}{1.3ex}
\begin{tabular}{l c*{14} l}
Field     
     & 
     \rot{AP-CSN} &
\rot{Change Sequence} &
\rot{Channel Usage} &
\rot{Cluster Probe} &
\rot{DSSS Parameters} &
\rot{Est. Service Param. Inb.} &
\rot{Est. Service Param. Outb.} &
\rot{Extended Request} &
\rot{FILS Request Parameters} &
\rot{Mesh ID} &
\rot{PV1 Probe Response Option} &
\rot{Request} &
\rot{S1G Relay Discovery} &
\rot{Vendor Specific Req.} &
\rot{SSID List} 
 \\
\toprule
In Probe Req. & \y & \y & \y & \y & \y & \y & \y & \y & \y & \y & \y & \y & \y & \y & \y \\
In Probe Resp. & \y & \y & \y & $\dagger$ & \y & \y & \y &  & (\y) & \y & $\dagger$ &  & \y & & $\dagger$ \\
In Assoc. Req. &&&&&&&&& (\y) && $\dagger$ && (\y) && $\dagger$\\
\bottomrule
  \multicolumn{16}{p{\linewidth}}{
    (\y): Fields further negotiating the specific tag are present\newline
    $\dagger$: Substituted by a different field\newline
  }
\end{tabular}
%  \multicolumn{16}{p{\linewidth}}
  %\begin{scriptsize}
%      {(\y): Fields further negotiating the specific tag are present\newline
%    $\dagger$: Substituted by a different field\newline}
  %\end{scriptsize}
\end{scriptsize}
\end{table}

When looking at association requests, 
%instead of containing e.g. FILS request parameters and S1G Relay Discovery, which are present in probe requests,
they contain elements negotiating FILS session activation and the use of S1G Relays. 
Therefore, their use must have been declared elsewhere when using passive discovery. An investigation of probe \textit{responses} confirms this: of the 15 elements not present in association requests, 8 are transmitted in probe responses.
%DSSS Parameters, Channel Usage, Mesh ID, Estimated Service Parameters Inbound, AP-CSN, Change Sequence, S1G Relay Discovery and Estimated Service Parameters Outbound. 
In addition to 84 other fields, probe responses (and also beacons) contain a FILS Indication field, which explains why devices using passive discovery can be aware of FILS capabilities without having requested information on it.

The fields which are exclusively present in probe requests are Request, Extended Request, and Vendor Specific Request, SSID List, FILS Request Parameters, PV1 Probe Response Option and Cluster Probe. 
The Cluster Probe element as well as the FILS Request Parameters are covered instead by the Extended Cluster Report field, respective FILS indication in probe responses and subsequent FILS parameters in the association frame. The SSID List element bundles multiple SSIDs that the device can connect to. In practice, requesting multiple SSIDs is instead done by sending separate probe requests for each SSID. 
The PV1 Probe Response Option \cite[Tables 9-305 - 9-310]{WLAN-spec} field is a bitmap of capabilities and compatibility, containing requests to respond with the capabilities the AP has. Most fields that can be requested via this bitmap are optionally present in probe responses and association requests. While this appears to be a useful field, connection establishment is possible, as well as efficient in passive discovery, without either the PV1 Probe Response Option or the  Request, Extended Request, and Vendor Specific Request tags. Therefore, these fields appear to be unnecessary in the exchange of capabilities. 

In fact, to test the variability in probe responses with respect to the probe requests they respond to, we monitor  probe responses in three different settings: in response to (1) scripted probe requests as used in \cref{sec: generic probe requests}, containing only an empty SSID field and Supported Rates, (2) those of the Intel 8265, and (3) those of a Raspberry Pi, with the longest IE field of all tested devices. 
\begin{observ}
    RRegardless of the IE content of probe requests, the AP always responds with the same capabilities.
\end{observ}
To confirm this, we observe the probe responses of 10 different APs in the vicinity of our setup over an extended period of time, spanning probe responses to various passersby and a large range of devices, with the same result.
We  conjecture that even probe requests containing a reduced IE content will receive all necessary capabilities of the AP via its probe responses. Since all relevant information in probe requests are also exchanged via the combination of probe response and association request, and the IE  content does influence probe responses, we conjecture that reducing the IE content to the bare minimum is feasible, and would reduce complexity and remove redundancy from wireless communication.

To estimate the improvement that generic probe requests have on the anonymity of single users, we subsequently analyse and compare anonymity sets.

%In fact, if the IE content is reduced to the bare minimum during \textit{undirected} probing, and conventional probe requests during \textit{directed} probing, the improvements supposed to be caused by the use of the IE (but not apparent in our experiments in \cref{sec: ie size test}) would still be in place. 

%   don't rely on our suggested probing improvements, but instead take place once the probing and the subsequent MAC address switch resulted in successful connection establishment, and probing using our method is discontinued at this stage.  while maintaining a good connection establishment speed. 

\subsection{Impact on Privacy: Anonymity Set Determination}\label{Anonymity Set Determination}

An anonymity set defines the number of users sharing the same identifiers, thus making them indistinguishable from one another. 
To estimate the privacy gain of generic probe requests, we evaluate the anonymity set size with reduced IE content like  Vanhoef et al. in 2016~\cite{Vanhoef-RandomisationNotEnough}. 
We use a subset of the Sapienza dataset~\cite{sapienza}, containing probe requests recorded at a train station in 2013, containing 376117 probe requests. MAC address randomisation was first introduced in iOS 8 in 2014~\cite{MAR-ios}, and in Android 6.0~\cite{MAR-android} in 2015. The dataset therefore contains \textit{very few probe requests that employ some form of MAC address protection.} It is perfect for the anonymity set determination: To perform the same analysis on more recent data, we would first have to apply de-randomisation strategies to the data set, e.g. define an attack that maps probe requests to single devices. This has been done in several publications (cf. \cref{sec: Related Work}.1), but is out of scope for this paper.

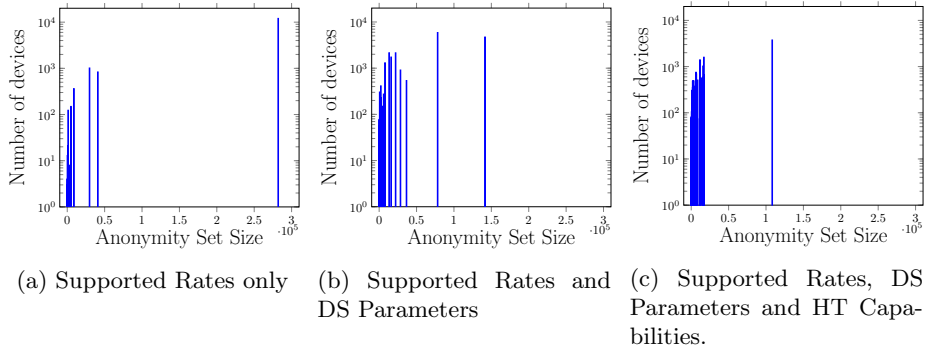
\begin{figure}[tb]
     \centering
  \begin{subfigure}[b]{0.32\textwidth}
    \centering 
  \resizebox{\textwidth}{!}{%
    \begin{tikzpicture}%[font=\large]
    %\tikzstyle{every node}=[font=\large]
            \begin{semilogyaxis}[xlabel={ Anonymity Set Size}, 
            ylabel={Number of devices}, 
            ybar, 
            xmin=0,
            xmax=300000,
            enlarge x limits={abs=10000},
            ymin=1,
            ymax=20000,
            bar width=0.5pt,
            label style={font=\LARGE},
            tick label style = {font = \normalsize},
]
                \addplot coordinates { (1, 1)( 5, 1)(5, 1) (5, 4) (11, 1) (56, 3) (93, 2)  (188, 1) (292, 13) ( 865, 10) (1232, 124) (1280, 21) (1485, 8) (2447, 8) (5009, 150) (8928, 364) (29807, 1018) (40916, 836) (282111, 12071)};
        \end{semilogyaxis}
    \end{tikzpicture}%
}
 \caption{Supported Rates only\vspace{0.75cm}}
    \label{fig:anonymity_set_sr}
\end{subfigure}
 \hfill
\begin{subfigure}[b]{0.32\textwidth}
    \centering 
    \resizebox{\textwidth}{!}{%
    \begin{tikzpicture}
            \begin{semilogyaxis}[xlabel={ Anonymity Set Size}, 
            ylabel={ Number of devices}, 
            ybar,
            xmin=0,
            xmax=300000, 
            enlarge x limits={abs=10000},
            ymin=1,
            ymax=20000,
            bar width=0.5pt,
            label style={font=\LARGE},
            tick label style = {font = \normalsize},
            ]
                \addplot coordinates {(1,1) (1,1) (1,1) (1,1) (1,1) (2,2) (4,3) (4,3) (4,3) (4,3) (4,2) (5,1) (5,1) (6,6) (6,4) (6,1) (9,1) (11,1) (14,10) (16,1) (17,16) (21,3) (56,3) (76,10) (77,1) (95,35) (106,20) (112,3) (124,15) (128,12) (173,9) (188,1) (201,76) (208,18) (281,27) (292,13) (298,57) (327,17) (363,39) (380,56) (430,78) (718,305) (865,10) (1199,92) (1232,124) (1280,21) (1485,8) (1969,78) (2133,230) (2256,414) (2447,8) (5009,150) (6010,275) (7858,1309) (13357,2155) (16038,1743) (21999,2149) (28502,913) (36555,540) (78201,5920) (141565,4736)  };
        \end{semilogyaxis}
    \end{tikzpicture}%
    }
 \caption{Supported Rates and DS Parameters\vspace{0.365cm}}
    \label{fig:anonymity_set_ch}
\end{subfigure}
\hfill
  \begin{subfigure}[b]{0.32\textwidth}
    \centering 
     \resizebox{\textwidth}{!}{%
    \begin{tikzpicture}
            \begin{semilogyaxis}[
            xlabel={Anonymity Set Size}, 
            ylabel={Number of devices}, 
            ybar, 
            xmin=0,
            xmax=300000,
            enlarge x limits={abs=10000},
            ymin=1,
            ymax=20000,
            bar width=0.5pt,
            label style={font=\LARGE},
            tick label style = {font = \normalsize},]
                \addplot coordinates {
(1,1) (1,1) (1,1) (1,1) (1,1) (1,1) (1,1) (1,1) (1,1) (1,1) (1,1) (1,1) (1,1) (1,1) (1,1) (1,1) (1,1) (1,1) (1,1) (1,1) (1,1) (1,1) (1,1) (1,1) (1,1) (1,1) (1,1) (1,1) (2,1) (2,2) (2,2) (2,1) (2,1) (2,2) (2,1) (2,2) (2,1) (2,1) (2,2) (2,1) (2,1) (2,1) (2,1) (2,2) (3,2) (3,1) (3,1) (3,2) (3,1) (3,2) (3,2) (3,1) (3,1) (3,3) (3,2) (3,1) (3,1) (3,1) (3,1) (3,2) (3,1) (4,2) (4,3) (4,1) (4,1) (4,1) (4,1) (4,1) (4,2) (4,1) (4,2) (5,1) (5,1) (5,1) (5,5) (5,3) (5,4) (5,1) (6,4) (6,1) (6,5) (6,1) (6,1) (6,2) (7,2) (7,6) (7,4) (8,5) (8,2) (8,1) (9,6) (9,4) (9,1) (9,2) (10,1) (11,1) (11,2) (11,1) (11,1) (12,2) (12,1) (12,4) (13,1) (14,5) (15,1) (15,3) (15,1) (16,1) (16,2) (17,3) (18,1) (18,2) (19,1) (21,3) (21,9) (21,1) (23,3) (23,1) (24,10) (25,4) (27,1) (30,6) (32,3) (33,2) (33,3) (33,18) (34,1) (35,3) (36,2) (37,1) (37,17) (39,12) (41,1) (44,1) (45,25) (45,11) (46,7) (46,1) (50,4) (50,12) (52,10) (52,8) (53,16) (53,1) (56,3) (57,2) (59,19) (60,1) (62,18) (63,3) (64,1) (65,12) (65,13) (66,1) (67,13) (77,10) (77,1) (82,47) (85,11) (85,4) (87,18) (87,1) (89,9) (89,5) (91,8) (91,19) (93,4) (99,17) (104,58) (111,2) (122,7) (123,12) (128,12) (128,55) (132,31) (144,22) (150,75) (155,11) (156,51) (160,6) (160,39) (161,1) (162,31) (167,5) (175,15) (176,60) (176,3) (188,1) (204,36) (214,38) (236,39) (237,44) (240,3) (242,23) (248,5) (255,6) (257,2) (275,77) (283,34) (289,37) (289,15) (292,13) (304,20) (311,34) (313,1) (354,76) (363,49) (377,82) (399,2) (425,5) (425,5) (467,69) (524,105) (528,17) (547,30) (581,40) (582,7) (607,4) (624,24) (662,77) (701,82) (714,77) (763,61) (777,168) (778,28) (783,125) (841,8) (865,10) (865,46) (900,1) (998,193) (1087,306) (1122,82) (1142,51) (1232,124) (1271,277) (1271,66) (1274,301) (1358,301) (1451,173) (1485,8) (1779,154) (1920,353) (1941,482) (2173,490) (2444,261) (2447,8) (2514,173) (2597,392) (2863,94) (3191,230) (3239,377) (3264,314) (3416,490) (3447,230) (3798,294) (3903,122) (4203,120) (4325,320) (4571,148) (6101,3) (6401,754) (6573,178) (6698,292) (6860,695) (8232,74) (8619,510) (11725,1403) (11748,81) (13304,568) (15881,1030) (15922,3) (16310,589) (16922,670) (16945,1598) (108324,3770) };
        \end{semilogyaxis}
    \end{tikzpicture}%
    }
 \caption{Supported Rates, DS Parameters and HT Capabilities.}
    \label{fig:anonymity_set_ht}
\end{subfigure}
\caption{The amount of devices contained in the same  anonymity set when the IE content is reduced to contain (a) only Supported Rates, (b) Supported Rates and DS Parameters, and (c) Supported Rates, DS Parameters and HT Capabilities.}
        \label{fig:anonymity_sets}
\end{figure}

Using the Wireshark filter \texttt{!(wlan.sa[0] \& 0x02)}, we determine which probe requests originate from globally unique MAC addresses via the U/L bit (cf.~\cref{sec: Background}), and find that 374736 are UAAs. These make up the pruned subset used for the subsequent evaluation. These probe requests originate from 14622 distinct MAC addresses. Since the dataset originates from before the introduction of MAC address randomisation, it is safe to assume that \textit{the number of distinct MAC addresses correlates with the number of devices}. 

The results of our evaluation can be seen in~\cref{fig:anonymity_set_sr}: The graph shows that  reduction of the IE to contain only Supported Rates results in 19 distinct anonymity sets. Around 17.45\% of the probe requests are in 18 smaller anonymity groups containing one to about 1000 devices. The remaining group unifies \num{82.55}\% of the devices in one large anonymity set. All of these devices share the same Supported Rates 2, 4, 11, and 22. 
\cref{fig:anonymity_set_ch} shows a reduction of the content of the IE to Supported Rates and DS Parameters. This reduction results in 61 anonymity sets, the largest of which now contains 5920 devices, which amounts to 40.49\%. In 225508 cases, the DS Parameters were not contained within the probe request. In the remaining cases, the DS Parameter set contained channels 1, 2, 11, or 12.

Figure \cref{fig:anonymity_set_ht} shows the anonymity set size in case the IE contains both Supported Rates, DS Parameters, and HT capabilities. The distribution is spread out significantly over 276 sets and the largest anonymity set contains \num{25.78}\%, or 3770, of the devices. %Their HT Capabilities have a tag length of 26 and contain only the information \texttt{0c 18 1b ff}, which amounts to a disabled SM Power Save, a maximum A-MSDU length of 7935 bytes, the capability of using DSSS/CCK in 40 MHz, A-MPDU Parameters, and the supported RX Modulation and Coding Scheme Set.
These results show that the less information is contained within probe requests, the larger the anonymity set most devices are contained in. It is therefore necessary to reduce and unify IE content as much as possible. To determine how much exactly, we regard prior attacks and the exact IE elements they use to fingerprint devices in the following.

\subsection{Impact on Security: Resistance to IE-Concerned Attacks}\label{sec: Resistance to IE-Concerned Attacks}

\begin{table}[tb]
\caption{A comparison of the fields used for IE fingerprinting. Publications that focused on certain fields with high entropy are highlighted in grey.}
\label{tab:ie-fingerprinting-attacks-modified}
%\begin{tabular}{lll}

    \scriptsize
    \newcommand\y{\checkmark}
      \setlength{\tabcolsep}{1.3ex}
      
 \begin{tabular}{l l*{16} l}
%  \begin{tabular}{l l l a l b a l*{9} l}
%    \begin{tabular}{c|c|c|c}
%\rowcolor{white\transparent{0.6}}
	Authors     
     & 
    \rot{Publication Year} &
    \rot{MAC Address} &
    \rot{SSID} &
    \rot{Supported Rates} &
    \rot{Ext. Supp. Rates} &
    \rot{DS Parameter Set} & %: Current Channel} &
    \rot{HT Capabilities } &
    \rot{VHT Capabilities} &
    \rot{Ext. Capabilities} &
    \rot{Vendor Specific} &
    \rot{all other fields } &
    \rot{IE Field Sequence}&
    \rot{BSS Membership} &
    \rot{Transm. Frequency} &
    \rot{Sequence No.} &
    \rot{RSS} &
%    \end{tabular}
 \\
%  \begin{tabular}{p{3cm} p{2ex}*2 a p{2ex} b a p{2ex}*{9}{l}}
\toprule
\rowcolor{gray!20}
Pang et al. \cite{pang80211User2007} & '07 & \y & \y & \cellcolor{lightblue!40}-& -& \cellcolor{mypurple!40}-& \cellcolor{lightseagreen!40}-& - & -& -&- & -& - & -& - & - \\ % MAC address, SSID & x \\
\rowcolor{gray!20}
Cunche et al. \cite{cuncheLinkingWirelessDevices2014} & '14 & \y & \y & \cellcolor{lightblue!40}-& -& \cellcolor{mypurple!40}- & \cellcolor{lightseagreen!40}-& - & -  & -&- & -& - & -& - & - \\ %  MAC address, SSID & x \\
\rowcolor{gray!20}
Freudiger \cite{TalkativePhones} & '15 &  \y & - & \cellcolor{lightblue!40}-& -& \cellcolor{mypurple!40}-& \cellcolor{lightseagreen!40}-& - & -& -&- & -& - & -& \y & - \\ % Sequence Number x \\

Vanhoef et al.\cite{Vanhoef-RandomisationNotEnough} &  '16 & \y  & \y  & \cellcolor{lightblue!20}\y& \y & \cellcolor{mypurple!20}\y &\cellcolor{lightseagreen!20}\y  & \y  & \y   & \y & \y  & \y & -  & -& - & - \\ %  All fields + IE sequence & c\\

Robyns et al. \cite{Robyns2017Noncooperative8M} & '17 &  \y  & \y  & \cellcolor{lightblue!20}\y&  \y &\cellcolor{mypurple!20}\y  & \cellcolor{lightseagreen!20}\y & \y  & \y   & \y & \y  & - & -  &  \y & - & - \\ % All fields depending on $\lambda$ and transmission frequency & "80.0 to 67.6 percent unique for 50 to 100 observed devices and 33.0 to 15.1 percent unique for 1,000 to 10,000 observed devices"\\
\rowcolor{gray!20}
Zhao et al. \cite{zhao2019localization} & '19 & \y & \y & \cellcolor{lightblue!40}-& -& \cellcolor{mypurple!40}- & \cellcolor{lightseagreen!40}-& - & -  & -&- & -& - & -& - & - \\ %  MAC address, SSID & x \\
\rowcolor{gray!20}
Dagelić et al. \cite{dagelic2019location} & '19 & \y & \y & \cellcolor{lightblue!40}-& -& \cellcolor{mypurple!40}- & \cellcolor{lightseagreen!40}-& - & -  & -&- & -& - & -& - & - \\ %  MAC address, SSID & x \\
Gu et al \cite{ProbeReqIdentification}&  '20 & - & - &\cellcolor{lightblue!20}\y& \y &  \cellcolor{mypurple!20}\y &  \cellcolor{lightseagreen!20}\y  & \y  & \y  & -  & \y &-  & - & - & -& - \\ % all fields except for MAC address, SSID, and vendor-specific values\\
\rowcolor{gray!20}
Uras et al. \cite{uras-ai}  & '20 & - & - & \cellcolor{lightblue!40}-& \y & \cellcolor{mypurple!40}\y & \cellcolor{lightseagreen!40}\y & \y &  \y & - & -& - & \y  & - & -& - \\ % Extended Supported Rates, BSS Membership, HT Capabilities, Extended Capabilities, DS Parameters, VHT Capabilities and two reserved tags & x \\
Tan et al. \cite{tan2021efficient} & '21 & \y  & \y  &\cellcolor{lightblue!20}\y& \y & \cellcolor{mypurple!20}\y &\cellcolor{lightseagreen!20}\y  & \y  & \y   & \y & \y  & - & -  & - & \y & \y \\ %  All fields + Sequence Number + RSS & discrimination accuracy (> 80\%), V-measure score (> 0.85). \\
\rowcolor{gray!20}
Pintor et al. \cite{pintor-ie-fingerprinting} & '22 & - & - & \cellcolor{lightblue!40}-& -& \cellcolor{mypurple!40}-& \cellcolor{lightseagreen!40}\y & - & \y & \y & -& - & - & - & - & -  \\ % Extended Capabilities, HT Capabilities, and vendor-specific values  \\
He et al. \cite{He2023SelfSupervisedAO} & '23 & \y  & \y  & \cellcolor{lightblue!20}\y& \y & \cellcolor{mypurple!20}\y &\cellcolor{lightseagreen!20}\y  & \y  & \y   & \y & \y  & \y & -  & \y & \y & \y \\ %  All fields + Sequence Number + RSS  + transmission time. \\
\bottomrule
\end{tabular}
\end{table}

To gauge the impact that a reduction of the IE content has on the security, a comparison of the fields used in IE-concerned attacks is performed in \cref{tab:ie-fingerprinting-attacks-modified}. The table highlights the fields particularly regarded in this publication, including the Supported Rates, the DS Parameter Set and the HT Capabilities in blue, purple and green. Additionally, the rows containing publications that selected specific fields due to their distinguishing features and high entropy are highlighted in grey. The conclusion that becomes apparent in this visual comparison is that \textit{no publication that selected fields depending on high entropy chose to include the Supported Rates}. The publications that included the Supported Rates incorporated them not due to high entropy, but to maximise the fingerprint.
This shows that the Supported Rates play only a minor role in identifying devices. 
This knowledge, in combination with our calculation of the anonymity set size of reduced IE content, show that \textit{removing all tags but the Supported Rates and the empty SSID field prevents existing attacks}. This way, users are protected from device tracking via the IE contained in their probe requests, while ensuring that probe responses can still be received.

%To further increase privacy, we propose to make probe requests even more generic and less distinguishable by employing one generic MAC address for all devices. This is necessary, since MAC address randomisation is, as of today, not standardised yet \cite{MAR-rfc}, which allows manufacturers to devise their own randomisation scheme. Some, in fact, use persistent OUIs and randomise only the last 3 bytes of the MAC address \cite{Vanhoef-RandomisationNotEnough, martin2017study}. This can allow an attacker to infer information on the device from the randomised MAC address. %Additionally, randomising sequence numbers within bursts as well would impede the distinction between devices even more.  
%Our suggested mitigation in the shape of the use of generic addresses is described and evaluated in the following. 

\section{Conclusion}\label{sec: Conclusion}

Reducing the content of probe requests to the minimum while enlarging the anonymity set to the maximum has been evaluated from different perspectives: Our experiments showed, that the only IE tags necessary for probe requests to receive probe responses are the Supported Rates and  SSID, both of which can be empty. 
We argue to unify the content of probe requests by transmitting only Supported Rates and the SSID. We furthermore showed that devices with probe requests containing the largest IE set do not necessarily correspond to the fastest connection establishment.
On the other hand, while the very large size of \textit{directed} probe requests sent by the Intel 8265 and its fast connection establishment do not necessarily have to correlate, it still poses the question whether large IEs in \textit{undirected} probe requests serve a purpose at all, or whether a combination of generalised and very slim undirected probe requests and, upon identifying a known network, sending information-rich directed probe requests might improve both the privacy, as well as the connection establishment speed.

We argue that other factors than the size of the IE must be the cause for the efficiency of the connection establishment, and that for the sake of increased user privacy, reducing the IE content in \textit{undirected} probe requests to the minimum is a valid option. To evaluate the anonymity provided by such minimisation, we calculate the anonymity set size of the Sapienza train station data set and find out, that reducing IE content to contain only Supported Rates allows for 82.55\% of the devices to be contained in the same anonymity set. This is a significant discovery that would protect a large number of users with very little effort.

Altogether, generic probe requests as proposed in this paper inhibit attacks targeting the IE by reducing its content to the bare minimum and thereby protect user privacy. %devices from correlation attacks on the IE content. The vast majority of attacks on probe requests choose this vantage (cf. \cref{sec: Related Work}), attempting to correlate flows using IE content. 
%Since our approach , % and additionally using one generic address for all probe requests, 
%it can successfully inhibit such attacks. 

%\todo{final pass -- check for consistency: British spelling, etc.}%

\bibliographystyle{splncs04}
\bibliography{sources}

\end{document}